\documentstyle[sprocl]{article}

\def\be{\begin{equation}}
\def\ee{\end{equation}}
\def\bea{\begin{eqnarray}}
\def\eea{\end{eqnarray}}

\begin{document}

\title{Parity and Large Gauge Invariance in Thermal  QED$_3$}

\author{D. Seminara}
\address{Department of  Physics, Brandeis University,
                        Waltham, MA 02454, USA}


\maketitle\abstracts{ We settle the ``apparent'' paradox present in
thermal QED$_3$ that the perturbative series is not invariant,
as manifested by the temperature dependence of the induced  
Chern-Simons term, by showing that large (unlike small) transformations 
and hence their Ward identities, are  not perturbative
order-preserving. Instead the thermal effective gauge field actions 
induced by charged  fermions  in $QED_3$ can be  made invariant  under 
both small  and  large gauge transformations by suitable  
regularization of  the  Dirac operator  determinant, at the usual price 
of parity anomalies. Our result is illustrated by a concrete example.}

Three-dimensional gauge theories are of physical 
interest in the condensed  matter context \cite{CondMatt}, but
display  special features requiring  understanding different from 
their four dimensional counterparts. In particular, we will
be concerned with the complex of problems associated  with the presence of
Chern-Simons  (CS) terms \cite{CS}, the necessary quantization of their
coefficients \cite{CS,Quant}
in the action stemmming from the possibility of making homotopically
nontrivial
{\it ``large"} gauge transformations, and the
effect
of quantum loop corrections on this sector \cite{loop}.
While large transformations  are
always relevant in the nonabelian case, they also come into play in the
physically most interesting case of $QED_3$ at finite temperatures where
the compactified euclidean time/temperature provides a nontrivial, 
$S^1$, geometry.
These exotic features have been the subject of a large literature
\cite{paradox}, as they seemingly lead to a paradox: 
on the one hand, large gauge invariance appears to require
quantization of the CS term's coefficient; on the other, matter loop
contributions to the effective gauge field action at finite temperatures
yield a perturbative expansion in which it acquires temperature-dependent, 
hence non-quantized, coefficients that seems to signal a gauge anomaly. This 
is particularly puzzling  since both the matter action and the process of
integrating out its excitations should be intrinsically gauge invariant.
Here, summarizing our previous results\cite{PRL,PRD}, we will show
that the effective action is indeed invariant under both
small and large transformations using classic results on elliptic 
operators\cite{Gilkey}
that allow a clear definition of  the Dirac operator's functional determinant 
by means of $\zeta-$function regularization. Instead, we will see that 
it is the perturbative
expansion that is non-invariant  because large transformations
necessarily introduce  non-analytic dependence on the charge so that
expansion in $e^2$ and large gauge invariance are mutually incompatible: the
induced Chern-Simons term's non-invariance is precisely 
compensated by further
non-local contributions in the effective action. We will also note
the necessary clash between gauge invariance and parity conservation,
similar to that in the familiar axial anomaly in even dimensions. 
All these features  are illustrated 
by explicit consideration of a    non-trivial configuration  
 that enables us  to ``parametrize'' the       CS     aspects.

Let us begin with the peculiar properties of large   gauge
transformations that invalidate the usual Ward identity 
consistency. Restoring explicit 
dependence on $e$, we have
$A_\mu\to A_\mu+e^{-1}\partial_\mu f$. Normally, we can merely
redefine $\tilde f=e^{-1} f$. This is also true at
finite temperature for the small gauge transformations since $f$
is only required to be periodic in Euclidean time $\beta=(\kappa T)^{-1}$.
Thus  a perturbative expansion will be small gauge
invariant  order by order. But for large ones, the periodicity condition becomes
$
f(0,{\bf r})=f(\beta,{\bf r})+2\pi i n$, with  $n\in Z\!\!\! Z$,
and a rescaling will merely hide the $e^{-1}$ factor in the boundary
conditions. This intrinsic dependence means  that only the
{\it full} effective action,
but not its individual expansion terms (including CS parts !) remains
invariant. 
We are therefore driven to a careful treatment of the
induced effective action $\Gamma[A]$ resulting from integrating out
the charged matter, for us massive fermions, according to the usual
relation
$
\exp\left(-\Gamma[A]\right)=\det(i D\!\!\!\!/+i m)
$
where $D_\mu$ is the $U(1)$ covariant derivative.
The extension to $N$ flavors and to the non-abelian case 
will be seen to be straightforward.
Our 3-space has $S^1({\rm time})\times \Sigma$ topology, 
$\Sigma$ being a compact Riemann 2-surface.
 We work with a
finite 2-volume in order to avoid infrared divergences  associated with
the continuous spectrum in an open space. Before proceeding, let us see
how assuming gauge invariance   constrains the form of the determinant. 
[To avoid irrelevant spatial homotopies, we shall 
here take $\Sigma$ to be the
sphere.] Because of the existence of the non-trivial $S^1$ cycle we can
construct (besides $F_{\mu\nu}$) the gauge invariant holonomy 
 $\Omega({\bf r})
\equiv\exp\left (i\int^\beta_0 A_0(t^{'},{\bf r})~dt^{'}\right )$. 
$\Omega$ is not a completely independent variable,
as  part of the information carried by it is already present in
$F_{\mu\nu}$: it satisfies the constraint ${\bf \nabla} \Omega=i
\Omega\int^\beta_0~{\bf E}(t^{'},{\bf r})~dt^{'}$,  implying
that $\Omega$ has the form $\Omega=\exp\left(2\pi i a\right)
\Omega_0({\bf E})$, where $\Omega_0({\bf E})$ is a nonlocal functional depending only
on ${\bf E}$ and on the geometry of $S^2$. The new information is encoded entirely
in the constant $a$, the  flat connection. [For  example, the
non trivial behavior of $A_0$ under large gauge transformation is inherited
by $a$: $a\to a+1$.] Therefore the determinant
can be considered as a
function(al) of $F_{\mu\nu}$ and $a$ alone. Large gauge invariance 
implies the separate Ward identity $e^{-\Gamma(a+1, F)}=
e^{-\Gamma(a,F)}$,
namely periodicity. Then Fourier-expanding and factorizing out the parity 
anomaly contribution, we obtain
\begin{equation}
\label{coppola}
e^{-\Gamma(F,a)}=
e^{iS_{CS}}\sum_{k=0}^{\infty}\Gamma^{(1)}_k(F_{\mu\nu})
e^{\pi(2 k-\Phi(F)) a}+\Gamma^{(2)}_k(F_{\mu\nu}) e^{-\pi(2 k-\Phi(F))},
\end{equation}
where $\Phi(F)$ 
is the electromagnetic flux through $S^2$ and $S_{CS}$ is  the abelian
CS action. [The correct form for the abelian CS action, when $A$
carries a nontrivial flux, hides some subletities that have been 
reviewed in\cite{PRD,CMP}.] As we shall see, the structure exhibited in  
(\ref{coppola}) will be explicitly realized in our example.

We now return to  the definition of the effective
action. Within our framework, the Dirac operator is a well-defined elliptic
operator \cite{Gilkey} whose determinant can be rigorously 
specified. The 
$\zeta-$function regularization \cite{Hawking}  defines the formal 
product of all the eigenvalues  $\lambda_n$ as
\begin{equation}
\label{3}
\det i\left(D\!\!\!\!/+m\right)=\Pi \lambda_n
\equiv\exp\left(-\zeta^\prime(0)\right),
\ \ \ \ \ \zeta(s)\equiv\sum  (\lambda_n)^{-s}
\end{equation}
with implicit repetition over degenerate eigenvalues. For $s>3$ in $D=3$
\cite{Gilkey}, the above series converges and its analytic extension 
defines a meromorphic function with only simple poles. It is regular
at $s=0$, thereby assuring the meaningfulness of (\ref{3}). A careful
definition of $\lambda_n^{-s}$ is required to avoid ambiguities. We take
it to be $\exp\left (-s \log\lambda_n\right)$ where the cut is chosen to
be over the positive real axis, $0\le\arg\lambda_n<2\pi$, enabling us to
rewrite $\zeta(s)$ in the more convenient form
\begin{equation}
\label{4}
\zeta(s)=\sum_{{\rm Re}~\lambda_n >0}(\lambda_n)^{-s}+ \exp(  -i\pi~s)
\sum_{{\rm Re}~\lambda_n <0}(-\lambda_n)^{-s}.
\end{equation}
Changing the cut only alters the determinant if it intersects the line
${\rm Im} z=m$, in which case the only relevant difference is the 
sign of the exponential in (\ref{4}).   This alternative choice does not affect 
gauge invariance, but
does change the sign of the parity anomaly terms in $\Gamma[A]$. 
Once the determinant
of the Dirac operator has been regularized, its full gauge invariance 
reduces to 
that of its eigenvalue spectrum. But small transformations do not affect the
$\lambda_n$ at all, while the large ones merely permute them, as  in 
usual illustrations of index theorems \cite{Gilkey};   every
well-defined symmetric function of the spectrum, such as $\zeta(s)$ and 
hence $\Gamma[A]$, is unchanged.

The price paid for preserving gauge invariance is (as usual !) an intrinsic
parity anomaly, {\it i.e.}, one present even in the limit when the 
explicitly parity violating fermion mass term is absent. [ That the parity can 
be  sacrificed  for gauge  was effectively noted in \cite{Red}.]
Under $P$,
 $\lambda_n\!\!\to\!\! -\lambda^*_n$  so that $\zeta^P(s)\ne\zeta(s)$. It is easy
to express the parity violating part $\Gamma^{(PV)}[A]=
1/2(\zeta^\prime(0)-\zeta^{\prime P}(0))$ explicitly in terms of the 
$\eta$-function in this limit. 
Here, $\zeta^{PV}\!\!\equiv
\zeta^P-\zeta$ is
\begin{equation}
\zeta^{PV}(s)=(1-e^{-i\pi s})\left(\sum_{\lambda_n >0}
(\lambda_n)^{-s}-\sum_{\lambda_n <0}(-\lambda_n)^{-s}\right)\equiv
(1-e^{-i\pi s})\eta(s),
\end{equation}
so that  $\Gamma^{(PV)}[A]=i\pi/2 \eta(0)$. At $m=0$, the continuous 
part of $\eta(0)$ is given in closed form by the CS action \cite{Gilkey}; 
being local means it can be removed by a different choice of regularization.
For $m\ne0$ an expansion in powers  of the mass can be presented
\begin{equation}
\label{potta2}
\Gamma^{(PV)}(A)=
\frac{1}{2}\left.\frac{d}{d~s}( \zeta^{PV}(s))\right|_{s=0}=
i\frac{\pi}{2} \eta(0)-
i \sum_{k=0}^\infty  (-1)^k\frac{m^{(2 k+1)}}{2k+1}
\eta (2 k+1),
\end{equation}
while the analogous expansion for the parity-conserving part involves
even powers of the mass. A detailed discussion of  mass 
expansions and the relevant references can be found in \cite{PRD}.

A simple, but  realistic, $(2+1)$ example\cite{PRL,PRD}, that
illustrates the
previous analysis, is the $U(1)$ field   
\begin{equation}
\label{7}
A_\mu (t,{\bf r})\equiv\left(\frac{2\pi}{\beta} a , {\bf A}({\bf r})\right ),
\end{equation}
where $a$ is a flat connection along $S^1$. ${\bf A}$ lives on
$\Sigma$, with non-vanishing, necessarily integer, flux $\Phi(F)=
n.$ We concentrate on large
transformations  $a\to a +1$, although in
higher genus $\Sigma$ one could also have large trasformations affecting
${\bf A}$. [ There is an intriguing $(0+1)$ antecedent of this example
\cite{Dunne,PRD,arg}.] 
Because of the time independence, we have a tractable eigenvalue 
equation for $\lambda_n$. After some work, it follows that the effective 
action factorizes into contributions:  one depending only on 
the holonomy  $\exp(2\pi ia)$
and the other    on the value of       ${\bf A}$ on 
$\Sigma$,
\begin{eqnarray}
\label{bilba}
&& e^{ -\Gamma(A)}=
\left [e^{  -\beta m+2\pi i a}+1\right]^{\nu_+}
\left [e^{  -\beta m-2\pi i a}+1\right]^{\nu_-}\\
&&\left |\prod_{\mu_k}\left(1+e^{       -\beta\sqrt{\mu^2_k+m^2}
+2\pi  i a }      \right)\right |^2\ 
e^{       2\pi ~\zeta_{\frac{\beta^2}{4\pi^2}
({/\!\!\!\!{\hat D}}^2+m^2)}(-1/2) - (\nu_+ +\nu_-) m \beta}       .
\nonumber
\end{eqnarray}
Here 
${/\!\!\!\!{\hat D}}$ is the reduced Dirac operator on $\Sigma$, 
$\mu_k$ its nonvanishing eigenvalues. The number  of 
positive/negative  chiral zero-modes $v_{\pm}$ of ${/\!\!\!\!{\hat D}}$ 
is represented by $\nu_{\pm}$, with the conventions 
$(\gamma_5\mp 1)v_{\pm}=0$, and the (parity odd) flux is just
$\nu_- -\nu_+$. 
That the infinite product in (8)
is convergent  follows from the fact that\cite{Gilkey} 
$\mu_k\simeq c\sqrt{|k|}$. The  invariance of (\ref{bilba}) under
$a\to a+1$  is 
manifest and its structure is consistent with (\ref{coppola}). 
It is clear that a perturbative (i.e., in power of $a$)
expansion of (\ref{bilba}) loses
periodicity in $a$ and hence does not see large invariance order by 
order. For example the Chern-Simons term  ( $S_{CS}=\pi a n$) has a 
coefficient $1-\tanh\left(\frac{\beta m}{2}\right )$. 
The usually quoted coefficient omits the $1$ that represents the 
intrinsic parity-anomaly price of our gauge-invariant regularization and hence
persists at $m=0$. There is actually an ambiguity in its sign (reflecting
the choice of cut in (3)), also present in other regularizations,
for example through the factor $\lim_{M\to\pm\infty} {\rm sign}(M)$ in Pauli-Villars.
 

In conclusion, we have shown that the apparent large gauge anomalies resulting 
from a perturbative expansion of the full effective action are due to
the more complicated (order-violating) nature of the Ward identities when a non-trivial
homotopy is  present, the action itself being fully gauge invariant with 
suitable regularization, one that necessarily entails parity
anomalies. 

It is a pleasure to thank my collaborators S. Deser and L. Griguolo.
This work was supported by NSF grant PHY-9315811.

\end{document}